# Cosmology from Transactional Entropic Gravity: A Concise Review

R. E. Kastner and Andreas Schlatter

March 30, 2026

Abstract: This is a review of key aspects of a model presented at the Lake Como School: Dark Matter, Dark Energy, and the Cosmological Tensions, June, 2025. The associated publication can be found at: A Schlatter and R E Kastner 2023 *J. Phys. Commun.* **7** 065009

## 1. Introduction

Based on the Relativistic Transactional Theory of Quantum Mechanics [1] the authors have developed a theory of gravity originating from the energy and entropy consequences, induced by transactions [2,3]. Curved spacetime and the Einstein equations are a derivable from transactions and the fact that the exchange of momentum in a transaction has a repulsive effect on matter enters mathematically into the Einstein equation as a constant which is proportionate to the transaction-density. This constant plays the role oft the Cosmological Constant $\Lambda$. It turns out to be indirectly proportionate to the squared radius of the universe [4]. By the exchange of light-particles there exists a natural clock in the cosmos which links the time-parameter based on any internal periodic process of a material system to spatial distance between systems. The proportionality-factor, namely the speed of light in vacuum, *c,* is invariant. Transactions create a special type of light-clocks, namely thermal clocks, in which the temperature equals the Davies-Unruh temperature associated to acceleration. By the invariance of the speed of light, these clocks must all march in step. These facts imply that a stationary observer in a region of the expanding comsos with a central mass will attribute to any test-mass at astronomical distance an acceleration which matches the MOND acceleration [5]. This way the effects of a cosmological constant as well as of dark matter in association with acceleration of distant masses are natural consequences of transactional entropic gravity.

## 2. Transactional Quantum Mechanics

The Relativistic Transactional Interpretation of quantum physics, (RTI) suggests that the spacetime manifold does not exist as a primary ontological structure, but supervenes on specific sorts of interactions among quantum systems--i.e. *transactions*, which will be briefly reviewed here. Quantitative details of this formulation are given in [1]. RTI predicts the emergence of spacetime events from a more fundamental *quantum substratum*, such that an actualized transaction gives rise to an emission event, an absorption event, and the invariant null interval connecting them, which is constituted by the real, transferred photon. Thus, this is a "growing universe'' type of picture.[1] In what follows, we show how the concept of a transaction, as elaborated in RTI, constitutes the crucial "missing link'' between the quantum level and the emergent spacetime level that allows for a fully consistent and seamless formulation of entropic gravity.

RTI is based on the fully quantum form of the direct-action or "absorber'' theory of fields

---

[1] In this picture the "Big Bang" was simply the initial emergence of the structured set of events that we call "spacetime", and which continues to grow.

([1], *Chapter* 5). In this picture, unitary quantum processes and interactions include the mediation of force through direct connections via the time-symmetric propagator, which is oblivious to temporal orientation. Accordingly, such interactions are not spacetime processes, but act "behind the scenes'', as physical possibilities (cf. Kastner, Kauffman and Epperson [6]), to steer the probabilities for the actualizations of spacetime events (the latter involving a non-unitary process to be discussed further below). The systems involved, such as elementary particles and atoms, are subject to quantum entanglement and accordingly are described by Hilbert Space states. As elements of a vector space of $3N, N\epsilon\mathbb{N}$, spatial dimensions with complex amplitudes, such states (and accordingly the entities they describe) are clearly not commensurable with the structure of spacetime. Thus, the domain of the systems and their interactions cannot be 3+1 spacetime and accordingly, as mentioned above, we identify it as an extra-spatiotemporal quantum substratum (QS). This is actually consistent with standard general relativity, since it is well known that the curvature associated with any mass source is nonvanishing. Were the mass source within the spacetime construct, the curvature would have to locally vanish at the mass.[2] Another type of unitary process in the QS is the Schrödinger evolution of fermionic matter, such as electrons, or bound states such as atoms. One can attribute an internal "clock'' to such systems, e.g. corresponding to their spin [7], which serves to define their proper time. Actualizations of material systems define inertial frames for the emergent spacetime, though they themselves are not features of spacetime.

One requires a non-unitary interaction called a "transaction'' in order to generate spacetime events and their connective structure. The non-unitarity is inherent in the direct-action picture of fields (but is obscured in the standard formulation), and can be quantified in terms of decay rates ([1], *Chapter* 5). The non-unitary interaction is triggered by what is called "absorber response'' in the non-relativistic form of the theory, but at the relativistic level (RTI) is actually a mutual interaction among emitters and absorbers, which is called the "NU-interaction'' (for "non-unitary interaction"). A basic requirement for the NU-interaction is the satisfaction of conservation laws. It is only through the NU-interaction that a real, on-shell photon can be created ([1], *Chapter* 5). Once a real photon is created, energy is transferred from the emitter to one of the responding absorbers in a radiative process. This process establishes a "link'' of spacetime constituted by the emission event, the absorption event, and the transferred photon. These links are invariant null intervals. Thus, spacetime is constituted solely by invariant events and their invariant null connections.

As mentioned above, the events are not to be identified with the material systems (e.g., atoms) giving rise to them. That is, the systems themselves are never part of the spacetime manifold nor are they within it. It is only their *activities*--the events--that are elements of the spacetime manifold, as are the photonic connections between them, which establish the metric structure of the manifold. In common parlance one refers to a spacetime point being "occupied'' by matter, but in our picture that is really a shorthand for the idea that a material system has become associated with that point via an actualized event. (Of course, given the spreading of wave packets, such a system will not remain indefinitely localized). In the sequel we show that the process of transactions produces gravity and some of its consequences as an entropic force [2,3].

**3. Gravity**

Assume there is a transaction between two physical systems at relative rest through the exchange of a real (on-shell) photon. A transaction breaks the unitary symmetry of the interaction and localizes absorber and emitter. If it takes the photon in the rest frame of the absorber, say, a time interval $\Delta t_R = \frac{R}{c}$ from the emitter to the absorber, then, next to the time interval $\Delta t_R$, there is the spatial distance $R > 0$ being generated by the process. For symmetry reasons[3], the possible locations of the absorber lie equiprobably on the sphere $S_R$. We can define the total entropy of the sphere $S(S_R)$ around the emitter by:

$$S(S_R) = k_B I(S_R) = \frac{k_B}{2} N_R, \tag{1}$$

where $N_R$ is just the number of bits on the sphere. For this number we set, with $l_P = \sqrt{\frac{G\hbar}{c^3}}$ and $A_R = 4\pi R^2$:

$$N_R = \frac{A_R}{l_P^2} = 4\pi \frac{R^2 c^3}{G\hbar}.\,^{4} \tag{2}$$

---

[2] Thus, the common notion that "everything physical exists in spacetime" is an uncritical folk belief that should be discarded.

So we get:

$$S(S_R) = k_B \frac{A_R}{2l_P^2}.\,^{5} \tag{3}$$

Expression (1) is the sum of all point-entropies $S(x), x \epsilon S_R$, which in turn are calculated over all possible local probabilities, $0 \leq p_x \leq 1, x \epsilon S_R$ [3]. So $S(S_R)$ can be considered to be the entropy contained in all possible absorptions on the sphere $S_R$.

Let us assume that in a concrete physical situation the material fields defined over the ball $B_R$ have the total mass $M$. We then know the total energy to be $E_{tot} = Mc^2$ and derive by the thermodynamic equivalence principle from the thermodynamic relation $E = S \cdot T$ the existence of a formal temperature, $T = T(M,R)$, satisfying:

$$Mc^2 = S(S_R) \cdot T(M,R) = k_B \frac{A_R}{2l_P^2} T(M,R). \tag{4}$$

With $g_R \stackrel{\text{def}}{=} \frac{GM}{R^2}$, weg get from (4) the expression for the Davies-Unruh temperature associated to acceleration $g_R$:

$$T(M,R) = \frac{\hbar M G}{2\pi c k_B R^2} = \frac{\hbar g_R}{2\pi c k_B}. \tag{5}$$

We are now ready for the main step.

### 3.1 Entropic force

In the following we reconstruct the argument in [8]. Assume that a particle of mass $m$ enters into a transaction with a system of total mass $M$, $m \ll M$ at relative rest, a process which localizes the particle at a point $x_0 \epsilon S_R$ for some $R > 0$. This process makes position-information available and consequently by the 2nd law, there is a (minimal) amount of entropy $\Delta \tilde{S}(x_0)$ added to the thermal environment on the surface. Since the Planck length is the smallest length possible, the entropy of the point-particle is $\tilde{S}(x_0) = 4\pi k_B I(x_0)$. Due to (3) we therefore have $\tilde{S}(x_0) = S(S_{l_P})$[6]. We further account for the fact that a particle is not really point-like, but still structureless, by using its (reduced) Compton radius $\lambda_C = \frac{\hbar}{mc}$ and postulate that the full information is part of the space-like surface $S_R$ only, if the particle is at a Compton-distance from it and that the information decreases linearly, if it is getting closer to the surface [8,9]:

$$I_{\Delta x}(x_0) = \frac{1}{2}\frac{mc}{\hbar} \cdot \Delta x, 0 \leq \Delta x \leq \lambda_C. \tag{6}$$

---
[3] Photons are emitted in all spatial directions with equal probability.

[4] Note that equation (2) might serve as a definition for the constant $G$ [8].

[5] We can multiply the information-entropy $I(x)$ by the "number" of bits on the sphere, because $I(x)$ is the average information over all partitions containing the point $x$ and is hence independent of a specific coverage $\mathcal{B}$ of the sphere. In order to count the number of bits on the sphere, $S_R$, we chose the local surface-measure $d\sigma(x) = l_P^{-2} d^2(x)$ and get $S(S_R) = \int_{S_R} k_B I(x) \, d\sigma(x) = k_B \frac{A_R}{2l_P^2}$.

Associated with the entropy difference there is by (5) a (minimal) amount of energy $E_{\Delta \tilde{S}(x_0)}$ given by:

$$E_{\Delta \tilde{S}(x_0)} = 4\pi k_B I_{\Delta x}(x_0) T(M, R). \tag{7}$$

The definition of energy as work, i.e. force along a path, leads to a corresponding force $F_G$ by:

$$E_{\Delta \tilde{S}(x_0)} = F_G \cdot \Delta x. \tag{8}$$

By (5) and (6) we get from (8):

$$F_G \cdot \Delta x = 2\pi k_B \frac{mc}{\hbar} \frac{\hbar MG}{2\pi c k_B R^2} \cdot \Delta x, \tag{9}$$

and finally the expression for $F_G$:

$$F_G = m \cdot g(M, R) = m \frac{GM}{R^2}. \tag{10}$$

The force $F_G$ is attractive, since the entropy-gradient $\Delta \tilde{S}(x_0)$ points to the surface $S_R$. The above derivation of $F_G$ has shown that the concept of spatial information and transactions leads, together with the thermodynamic equivalence principle (4), to the existence of gravity as an entropic force, emerging from the coming-into-being of events.

Once we consider that transactions possess by the exchanged photons naturally inbuilt clocks, which measure the rhythm of becoming by an invariant gauge $c$, the speed of light, the way is open to derive a metric structure of spacetime, locally governed by the Einstein equations. The derivation of Einstein's equation in this way is done in [2,3].

### 4. The Cosmological Constant

In transactional entropic gravity the energy of the transferred photons gauges the rhythm of becoming as it defines the period of natural light-clocks. This ultimately leads to Einstein's equation [3]. The three-momenta of the transferred photons enter this equation in form of a cosmological constant, $\Lambda$, as we will now show [4,10]. The idea is intuitively plausible, since the momenta of the photons exercise a repulsive pressure on the material systems involved in the transactions. Concretely, there arises pressure from photon three-momenta, emitted equiprobably in the spatial directions, which defines at a given point the Laue-scalar $T$:

---

[6] Entropy is understood as unavailable position-information, which is calculated as an average over all possible systems. Transactions make this information available, a fact which must be compensated by the second law. In this sense "a particle adds one bit of information to the surface" [8].

$$T = \sum_{i=1}^{3} T_{ii} = \lim_{A_i \to 0} \sum_{i=1}^{3} \frac{F_i}{A_i} = \lim_{A_i \to 0} \sum_{i=1}^{3} \frac{1}{A_i} \frac{dp_i}{dt}. \tag{11}$$

Let $N_R(t)$ be the number of actualizations within (spatial) volume $V_R$ at a time $t$. We have with $x_0 = ct$, $N_R(t) = N_R\left(\frac{x_0}{c}\right) = \widetilde{N}_R(x_0)$ and with the de Broglie relation $|\vec{p}| = \frac{h}{R}$ [7] and $\delta(x_0) \stackrel{\text{def}}{=} \frac{\widetilde{N}_R(x_0)}{V_R}$:

$$T = -3 \frac{dN_R(t)}{dt} \cdot \frac{1}{A_R} \cdot \frac{h}{R} = -3 \frac{c \cdot h}{3} \cdot \frac{d\widetilde{N}_R(x_0)}{dx_0 V_R} = -c \cdot h \cdot \frac{d}{dx_0} \delta(x_0). \tag{12}$$

The negative sign indicates the repulsive effect and the function $\delta(x_0) = \frac{\widetilde{N}_R(x_0)}{V_R}$ denotes the number of transactional events per spatial volume at time $x_0$, in short: the transaction-density. The term $\frac{d\delta(x_0)}{dx_0}$ is therefore the change-rate of the transaction-density. Hence, there holds to first order: $\delta(x_0 + \Delta x_0) = \delta(x_0) + \frac{d\delta(x_0)}{dx_0} \cdot \Delta x_0$. In (12) we assumed that $\delta(x_0)$ is constant over space, which also amounts to the homogoneity and isotropy of space with respect to transactional events. Equation (12) also tacitly assumes that $\delta(x_0)$ is a differentiable function in $x_0$. This is an assumption which cannot hold in quantum-mechanics, since quantum-events represent discrete sets and are not deterministic, but obey a random-process. As it happens, the emission-absorption of photons is a decay-process and the associated event-number variable is (for large numbers) naturally modeled by a Poisson-distribution with constant average transaction-density rate $\varrho_\gamma$. Hence, in analogy to the above terminology we have for the average transaction-density $\bar{\delta}(x_0)$ and for $\Delta x_0 > 0$:

$$\bar{\delta}(x_0 + \Delta x_0) = \bar{\delta}(x_0) + \varrho_\gamma \cdot \Delta x_0. \tag{13}$$

So, by (13) we can define, in analogy to (12):

$$T_\gamma = -3 \frac{c \cdot h}{3} \cdot \frac{\Delta \bar{\delta}(x_0)}{\Delta x_0} = -c \cdot h \cdot \varrho_\gamma. \tag{14}$$

So far, we have worked in a specific frame and found a specific type of stochastic process for the transaction-density. A Lorentz-invariant[8] distribution (and arguably the only one) for a homogeneous and insotropic (i.e. uniform) spreading of events in spacetime, such that the number of events is proportionate to the spacetime volume, just happens to be a Poisson-distribution with constant average density [11]. This means that independent of a reference frame, we can consider $\varrho_\gamma$ to be a four-density (and hence a scalar) and the number-variable obeys a Poisson-distibution with factor $\lambda = \varrho_\gamma V$ for any given four-volume $V$, guaranteeing the Lorenz-invariane of the "transactional sprinkling" in every sense [12][9].

Remember now the Einstein equation:

---

[7] We choose the minimal energy necessary to meaningfully localize a particle within distance $R$.
[8] Lorentz invariance has two meanings in this context. First, the invariance of the uniformity of the event-distribution per-se and second, the impossibility to choose a preferred time diection in any realization of a spreading.
[9] The constancy of the speed of light and hence the validity of the syncalibration of thermal clocks with light clocks is ultimately equivalent to the invariance of the number of events in a spacetime volume.

$$R_{\mu\nu} = \frac{8\pi G}{c^4}\left(T_{\mu\nu} - \frac{1}{2}T g_{\mu\nu}\right), 0 \leq \mu, \nu \leq 3. \tag{15}$$

Equation (14) is at first part of the energy-momentum of matter and radiation and hence leads on the right-hand side of Einstein's equation, with $l_P = \sqrt{\frac{G\hbar}{c^3}}$ denoting the Planck length, to the term:

$$\frac{4\pi G}{c^4} T_\gamma = -\frac{4\pi G \hbar}{c^3}\varrho_\gamma = -8\pi^2 l_P^2 \varrho_\gamma. \tag{16}$$

The right-hand side of equation (15), however, consists of the local (average) energy-momentum distribution of actualized matter and radiation fields, whereas expression (16) stems from single photons and is global, i.e. independent of any specific local arrangement. It hence represents a structural component of equation (15). Consequently, it belongs to the left hand side of the Einstein equation. Hence, we can interpret the expression in equation (16) as a cosmological constant:

$$\Lambda \stackrel{\text{def}}{=} 8\pi^2 l_P^2 \varrho_\gamma. \tag{17}$$

Einstein's equation thus takes on the form:

$$R_{\mu\nu} + \Lambda g_{\mu\nu} = \frac{8\pi G}{c^4}\left(T_{\mu\nu} - \frac{1}{2}T g_{\mu\nu}\right), 0 \leq \mu, \nu \leq 3. \tag{18}$$

### *4.1 The Size of $\Lambda$*

In order to investigate the properties of $\Lambda$ we work in a FLRW-spacetime, where there is a natural foliation. It is by equation (17) necessary to understand today's transaction-density rate $\varrho_\gamma$. Since $\varrho_\gamma$ is an average and the transaction-density at the origin can be set to zero, $\bar{\delta}(0) = 0$ [10], we choose the direct approach and estimate the spatial density of the expected number of transactions in today's causal universe and divide it by the age of the universe, in order to get the average transaction-density rate today. To make close calculations possible, we employ a simple model of an expanding, flat universe. Furthermore, we have theoretical [11] and observational evidence for a decreasing Hubble parameter $H(t)$ [13]. If $t_U$ denotes the age of the universe today, we set for today's Hubble parameter $H(t_U) \stackrel{\text{def}}{=} H_0$ [12] and use today's Hubble radius $R_{H_0} = \frac{c}{H_0}$ to express $t_U$, which is $t_U = \frac{1}{H_0}$. We further know from the Friedmann equation that for the expansion factor, $a(t)$, we have $\frac{\dot{a}(t)}{a(t)} = H(t)$. Hence, $a(t) = a_0 e^{\int_0^t H(\tau)d\tau}$, with $a(0) = a_0$ chosen such that $a(t_U) = 1$.[13] The causal universe at any time $t \leq t_U$ is bounded by the particle horizon, $R_P(t)$, which is defined by:

---

[10] There are no transactions yet and the volume is, due to the Planck length, never zero.
[11] Matter and radiation-densities decrease over time.
[12] Note that in this situation the first Friedmann equation implies $H(t)^2 \sim \frac{\Lambda(t)c^2}{3}$.
[13] Note that this implies $a_0 \ll 1$.

$$R_P(t) \stackrel{\text{def}}{=} a(t) \int_0^t \frac{c d\tau}{a(\tau)} = a(t) \int_0^t \frac{c d\tau}{a_0 e^{\int_0^\tau H(s) ds}}. \tag{19}$$

Since $H(t) \geq H_0$, there holds:

$$R_P(t) \leq a(t) \int_0^t \frac{c d\tau}{a_0 e^{H_0 \tau}} = \frac{a(t)}{a_0} R_{H_0} (1 - e^{-H_0 t}). \tag{20}$$

In particular, for today's particle horizon, $R_P(t_U)$, we have with $\varepsilon \stackrel{\text{def}}{=} (1 - e^{-1})$: [14]

$$R_P(t_U) \leq \frac{\varepsilon}{a_0} R_{H_0}. \tag{21}$$

We further remember that a single point "$x_0$" in space represents one bit of information with information content $I(x_0) = \frac{1}{2}$. Remembering the Planck length as a minimal length in nature, the total number of bits, $n_R$, on the surface of a ball of radius $R$ with surface-area $A_R$ therefore amounts to:

$$n_R = \frac{A_R}{2 l_P^2}. \tag{22}$$

By the holographic principle expression (22) represents the maximum information encoded within the ball $B_R$. Furthermore, the fine-structure constant, $\alpha^2$,[15] is the base-probability for a transaction to happen (at most reduced by a factor stemming from additional specific amplitudes) and we can reasonably assume that the average number of transactional events at any time $x_0$ is (maximally) proportionate to the available spatial information.[16] Hence, the (maximum) expected number of transactional events within a ball of radius $R$, $B_R$, is:

$$\bar{N}_R = n_R \cdot \alpha^2 = \frac{A_R \alpha^2}{2 l_P^2}. \tag{23}$$

Since the causal universe has been growing to today's particle horizon, $R_P(t_U)$, we can therefore assume that the total amount of transactional events today is encoded on the surface of radius $R_P(t_U)$. By equation (23) we get for the total expected number of transactional events, $\bar{N}_{R_P(t_U)}$:

$$\bar{N}_{R_P(t_U)} = \frac{4\pi R_P^2(t_U) \alpha^2}{2 l_P^2}. \tag{24}$$

---

[14] $\varepsilon \approx 0.632$.

[15] There holds with $q$ denoting the elementary electric charge and $\varepsilon_0$ the dielectrical constant: $\alpha^2 = \frac{q^2}{4\pi\varepsilon_0 \hbar c} \approx \frac{1}{137}$.

[16] This holds because emitting or absorbing material systems are much larger in size than the Planck-length.

To get the average transaction-density, $\bar{\delta}(t_U)$, we need to divide expression (24) by the volume of the causal universe to get:

$$\bar{\delta}(t_U) = \frac{\bar{N}_{R_P(t_U)}}{\frac{4}{3}\pi R_P^3(t_U)} = \frac{3\pi\alpha^2}{2l_P^2 R_P(t_U)}. \tag{25}$$

Since $\bar{\delta}(0) = 0$, we have together with equations (13) and (21) for the average transaction density-rate:

$$\varrho_\gamma = \frac{\bar{\delta}(t_U) - \bar{\delta}(0)}{ct_U} \leq \frac{\varepsilon\bar{\delta}(t_U)}{a_0 R_P(t_U)} = \frac{3\pi\varepsilon\alpha^2}{2a_0 l_P^2 R_P^2(t_U)}. \tag{26}$$

By setting $C_0 \stackrel{\text{def}}{=} \left(\frac{3\pi\varepsilon}{2a_0}\right)$, which is a dimensionless number, and by equation (17) we finally get for the cosmological constant:

$$\Lambda \leq 8\pi^2 C_0 \frac{\alpha^2}{R_P^2(t_U)}. \tag{27}$$

In equation (27) we directly recover the measured fact that $\Lambda \sim R_U^{-2}$.[17] It becomes clear that $\varrho_\gamma$ and therefore $\Lambda$ are constant only over limited time-scales but not so over cosmic time-scales and we should expect to measure a decrease in the accleration of the expansion of the cosmos - beyond the thinning out of matter and radiation density - proportionate to the inverse square of the size of the universe. By the same token, $\varrho_\gamma$ might vary over different spatial regions leading to inhomogeneities across the universe. Besides this, the key role of the fine-structure constant becomes clear. It is the basic governing factor of the expected number of transactional events in the universe and as such enters the formula for its expansion.

**5. Galaxy Rotation**

The observation that the rotation-velocity of the outer rims in spiral galaxies does not match the predictions of Newtonian gravity and rotates too rapidly, given the observed mass of these large objects, has lead to the hypothesis of the existence of dark matter. Various proposals as to its true nature have been made, but so far dark matter has remained elusive. Another approach to explain the data has been to suggest modifications to Newtonian gravity, prominent among which is MOND-theory [5]. We will show that the acceleration and velocity found in MOND can be deduced from the syncalibration of the thermal clocks of two different observers [3]. The starting points are again equation (3), which assigns to a sphere of radius $r > 0$ an entropy of:

$$S(r) = \frac{k_B A_r}{2l_P^2}, \tag{28}$$

(where, as before, $A_r \stackrel{\text{def}}{=} 4\pi r^2$) and the Davies-Unruh temperature, measured by an observer with acceleration $\kappa$:

$$T_\kappa = \frac{\hbar\kappa}{2\pi k_B c}. \tag{29}$$

---
[17] We have today $R_P(t_U) = 46.5 Gly \approx 4{,}2 \cdot 10^{26} m$.

Equations (28), (29) lead to the thermodynamic energy-expression:

$$E_\kappa(r) = S(r) \cdot T_\kappa. \tag{30}$$

These equations and their modifications will prove key, this time in a specific spacetime, namely the Schwarzschild-de Sitter solution of Einstein's equation.

### 5.1. Schwarzschild-de Sitter Spacetime

To represent spacetime with a cosmological constant and with the presence of a central mass $M$, the corresponding Schwarzschild radius of which is $R_S = \frac{2GM}{c^2}$, we use the static patch of the Schwarzschild-de Sitter solution of Einstein's equation [14]. Its line-element has the form:

$$ds^2 = f(r)c^2 dt^2 - \frac{1}{f(r)} dr^2 - r^2 d\Omega^2, \tag{31}$$
$$f(r) = \left(1 - \frac{r^2}{R_0^2} - \frac{R_S}{r}\right),$$

where $R_0 = \sqrt{\frac{3}{\Lambda}}$. Note that the metric (31) is static, although it describes the perspective of a non-inertial observer. In the framework of transactions it follows that there is no such thing as a pure vacuum spacetime, since a metric structure and a cosmological term only arise by transactions and the corresponding radiation, which involves actualized matter-fields. From a mathematical perspective, though, it is of course possible to consider metric (31) as a combination of a vacuum solution to Einstein's equation with metric factor:

$$\tilde{f}(r) = \left(1 - \frac{r^2}{R_0^2}\right), \tag{32}$$

called de Sitter space, adjusted by the potential of a gravitating mass $M$, $2\Phi(r) = \frac{2MG}{c^2 r}$. $R_0$ is the Hubble-horizon in de Sitter space. Note that the addition of a mass $M$ reduces the horizon $R_0$, $\tilde{f}(R_0) = 0$, to some smaller radius $\hat{R}_0 < R_0, f(\hat{R}_0) = 0$, a fact which we will use in the sequel.

In de Sitter space there are the following identities involving the acceleration of the horizon $a_\infty$ [3]:

$$a_\infty = cH_0 = \frac{c^2}{R_0} = c^2 \sqrt{\frac{\Lambda}{3}}, \tag{33}$$

where $H_0$ is the Hubble-constant [15]. We define $a_0 \stackrel{\text{def}}{=} \frac{a_\infty}{2}$ and for an interpolated acceleration $a(r)$ at radius $r < R_0$ we set, by chosing the simplest ansatz:

$$a(r) \stackrel{\text{def}}{=} \frac{r}{R_0} a_0. \tag{34}$$

Since $T_{a(r)} = \frac{r}{R_0} T_{a_0}$ and the surface element in the static metric of de Sitter space is also $d\sigma^2 = r^2 d\Omega^2$, we have:

$$E_{a(r)}(r) = S(r) \cdot T_{a(r)} = \frac{r}{R_0} S(r) \cdot T_{a_0}. \tag{35}$$

Hence, the factor $\frac{r}{R_0}$ can also be attributed to the entropy-expression $S(r) = \frac{k_B A_r}{2 l_P^2}$ in (35) in order to define the de Sitter entropy $S_{dS}(r)$:

$$S_{dS}(r) \stackrel{\text{def}}{=} \frac{r}{R_0} S(r). \,^{18} \tag{36}$$

### 5.2 Effective Entropy in Schwarzschild-de Sitter Spacetime

Since, as mentioned above, the mass-induced potential $2\Phi(r)$ in (30) reduces the horizon, $\hat{R}_0 < R_0$, it also reduces, by an argument of Verlinde [16] the corresponding entropy (35). We follow the line of thought in [16]. For $|\Phi(r)| \ll 1$ the horizon changes through addition of mass $M$ by a negative amount of:

$$\Delta R_0 = \Phi(R_0) \cdot R_0. \tag{37}$$

Hence, the horizon-entropy changes under $R_0 \to R_0 + \Delta R_0$ to first order by:

$$\Delta S_{dS}(R_0) = \Delta R_0 \frac{d}{dR_0}\left(\frac{k_B A_{R_0}}{2 l_P^2}\right) = -\frac{4\pi c k_B M R_0}{\hbar}. \tag{38}$$

In order to calculate first order entropy change in regions far away from the horizon, $\frac{r}{R_0} \ll 1$, we take in the presence of a mass $M$ the derivative of $S(r)$ with respect to the geodesic distance $ds = (1 + \Phi(r))^{-1} dr$, whereas in vacuum this distance is simply $ds = dr$. The first order difference of entropy in the two situations thus becomes:

$$\frac{d}{ds}\left(\frac{k_B A_r}{2 l_P^2}\right)\bigg|_{M=0}^{M \neq 0} = \Phi(r) \frac{d}{dr}\left(\frac{k_B A_r}{2 l_P^2}\right) = -\frac{4\pi c k_B M}{\hbar}. \tag{39}$$

The right hand side of (38) is the amount of entropy, which a mass $M$ takes away from the entropy of a spherical region of radius $r$. We define the infinitesimal change $\frac{dS_{dS}(r)}{dr}$ to be differentiation with respect to the coordinate "$r$" only, to get after integration:

---

[18] A physical reason for definition (36), using arguments from string theory, is given by Verlinde [16].

$$\Delta S_{dS}(r) = -\frac{4\pi c k_B M}{\hbar} r. \tag{40}$$

Multiplying equation (40) both in the nominator and denominator by $R_0$ and remembering (38) leads to:

$$\Delta S_{dS}(r) = \frac{r}{R_0} \Delta S_{dS}(R_0). \tag{41}$$

We now define the effective de Sitter entropy after addition of a mass $M$, $\overline{S}_{dS}(r)$, by:

$$\overline{S}_{dS}(r) \stackrel{\text{def}}{=} S_{dS}(r) + \Delta S_{dS}(r) = \frac{k_B r}{R_0}\left(\frac{A_r}{2l_P^2} - \frac{4\pi c M R_0}{\hbar}\right). \tag{42}$$

We are now ready to tackle the main goal, namely to calculate the effective acceleration $\overline{a}$ in Schwarzschild-de Sitter spacetime.

### 5.3 The Interpolation-Formula

The effective acceleration $\overline{a}$ in Schwarzschild-de Sitter spacetime describes the point of view of a non-inertial observer for whom a test mass is subject to a combined vacuum-gravity acceleration. At the same time we can also take the viewpoint of an observer co-moving with the vacuum expansion who observes a test mass only experiencing gravitative pull $g(r)$ [17,18,19]. Both accounts have to be consistent. In order to express the co-moving observer, we work in the FLRW-form of the static metric (31), the McVittie-solution to Einstein's equation, with line-element [18]:

$$ds^2 = \left(\frac{1-\mu}{1+\mu}\right)^2 c^2 d\bar{t}^2 - A(\bar{t}, \bar{r})(d\bar{r}^2 + \bar{r}^2 d\Omega^2), \tag{43}$$

where $\mu = \frac{GM}{2c^2 a(\bar{t})\bar{r}}$, $A(\bar{t},\bar{r}) = (1+\mu)^4 a^2(\bar{t})$, and $a(\bar{t}) = e^{H_0 \bar{t}}$. The radius $\bar{r}$ is an isotropic coordinate with corresponding transformation into the Schwarzschild radial coordinate $r$, given by $r = a(\bar{t})\bar{r}\left(1 + \frac{GM}{2c^2 a(\bar{t})\bar{r}}\right)^2$.

Let us define at this place the following quantities:

$$d\sigma \stackrel{\text{def}}{=} \sqrt{f(r)}cdt, \qquad d\tau \stackrel{\text{def}}{=} \left(\frac{1-\mu}{1+\mu}\right)cd\bar{t}. \tag{44}$$

In distant regions where gravitational acceleration is sufficiently small, we are approximately in a vacuum, i.e. de Sitter space, and there should hold (34):

$$\overline{a}(r) \sim a(r). \tag{45}$$

We can say much more about $\overline{a}$ by noticing that both observers can locally define thermal clocks. By the principle of syncalibration this leads in the range $r > r_0$, $r_0$ to be determined later, to the equality:

$$\left(T_{\overline{a}(r)}\overline{S}_{dS}(r)\right)^{-1} d\sigma = \left(T_{g(\bar{r})}S(\bar{r})\right)^{-1} d\tau. \tag{46}$$

Since we operate at astronomical distances, $R_S \ll r \ll R_0$[19] and $\bar{t} \gg 1$, we have $\mu \ll 1$ and $f(r) \approx 1$. Hence, $r \approx a(\bar{t})\bar{r}$, $d\sigma \approx cdt$ and $d\tau \approx cd\bar{t}$. Thus, equation (45) turns into [20]:

$$\left(T_{\overline{a}(r)}\overline{S}_{dS}(r)\right)^{-1} dt = \left(T_{g(r)}S(r)\right)^{-1} d\bar{t}. \tag{47}$$

In addition, the two time-parameters $t, \bar{t}$ are related by a Gullstrand-Painlevé transformation with [17]:

$$dt = d\bar{t} + \frac{r}{R_0}\left(1 - \frac{R_S}{r}\right)^{-\frac{1}{2}}\left(1 - \frac{R_S}{r} + \frac{r^2}{R_0^2}\right)^{-1}, \tag{48}$$

such that at astronomical distances $dt \approx d\bar{t}$ and hence:

$$T_{\overline{a}(r)}\overline{S}_{dS}(r) = T_{g(r)}S(r). \tag{49}$$

Equation (49) is the key relation and by (42) we get:

$$T_{\overline{a}(r)}\left(S_{dS}(r) + \Delta S_{dS}(r)\right) = T_{g(r)}S(r). \tag{50}$$

By applying (36), (40) and (45) we arrive at:

$$\frac{\overline{a}(r)^2}{a_0}S(r) - \overline{a}(r)|\Delta S_{dS}(r)| - g(r)S(r) = 0. \tag{51}$$

For fixed $r \geq r_0$ (51) is quadratic in $\overline{a}(r)$ with one positive solution:

---

[19] Note that the Schwarzschild radius also changes in Schwarzschild-de Sitter spacetime, like the cosmic horizon. For reference, though, we use $R_S$. At the same time we use functions of coordinates $r, \bar{r}$ which do not exactly represent the radial-distance in the respective metrics, but are proportionate to it. Hence terms like "far out" are well reflected in the magnitude of $r$ and $\bar{r}$.

[20] For $r = a(\bar{t})\bar{r}$ we have $T_{g(\bar{r})}S(\bar{r}) = T_{g(r)}S(r)$.

$$\overline{a}(r) = a_0 \frac{\left(\left(\frac{|\Delta S_{dS}(r)|}{S(r)}\right) + \sqrt{\left(\frac{|\Delta S_{dS}(r)|}{S(r)}\right)^2 + \frac{4g(r)}{a_0}}\right)}{2}. \quad (52)$$

The Newtonian potential can be written as $|\Phi(r)| = \frac{1}{2}\frac{|\Delta S_{dS}(r)|}{S(r)}$ and hence (52) turns into:

$$\overline{a}(r) = a_0 \frac{\left(2|\Phi(r)| + \sqrt{4|\Phi(r)|^2 + \frac{4|\Phi(r)|^2 c^4}{MGa_0}}\right)}{2}. \quad (53)$$

After factoring out, we finally get the interpolation-formula:

$$\overline{a}(r) = a_0|\Phi(r)|\left(1 + \sqrt{1 + \frac{c^4}{MGa_0}}\right). \quad (54)$$

### *5.4 Discussion of the Interpolation-Formula*

We already mentioned that equations (46) and following are valid for radii, $r \geq r_0$, bigger than some limit-radius $r_0$. The size of this limit-radius follows immediately from equation (42), because for (49) to make sense there must hold:

$$\overline{S}_{dS}(r) > 0. \quad (55)$$

This means by (42):

$$\frac{A_r}{2l_P^2} > \frac{4\pi cMR_0}{\hbar}, \quad (56)$$

and hence:

$$r > r_0 = \sqrt{\frac{MG}{a_0}}. \quad (57)$$

Below this limit $\overline{S}_{dS}(r)$ is exhausted, gravity dominates and there holds the Newtonian regime. We still need to check consistency at $r = r_0$. To do this, we make use of the fact that by (42) we have $S(r_0) = \Delta S_{dS}(R_0)$ and hence $|\Phi(r_0)| = \frac{r_0}{2R_0}$. Since $\frac{r_0}{R_0} a_0 \ll 1$[21], equation (54) turns by (33) into:

$$\overline{a}(r_0) \approx \frac{r_0}{2R_0} a_0 \sqrt{\frac{c^4}{MGa_0}} = \sqrt{\frac{c^4}{4R_0^2}} = a_0. \tag{58}$$

On the other hand, we get by a direct calculation:

$$g(r_0) = \frac{c^2 |\Phi(r_0)|}{r_0} = \frac{c^2}{2R_0} = a_0. \tag{59}$$

Equations (58) and (59) guarantee consistency of the regimes at $r = r_0$.

At the same time, since $|\Phi(r)|a_0 \ll 1$ for $r > r_0$, we can simplify equation (54) further to get:

$$\overline{a}(r) \approx |\Phi(r)|a_0 \sqrt{\frac{c^4}{MGa_0}} = \frac{\sqrt{MGa_0}}{r}. \tag{60}$$

The right hand side of (60) is exactly the expression suggested in the original MOND-theory [5] at accelerations below $a_0$. Via the relation $\overline{a}(r) = \frac{v^2(r)}{r}$ we finally get:

$$v^2 = \sqrt{MGa_0}, \tag{61}$$

which fits observed velocity-data in spiral galaxies [20]. Equation (61) establishes a Tully-Fisher- type relation, $v^\beta \sim M$, $\beta \epsilon \mathbb{R}$, between the baryonic mass $M$ of a galaxy and the rotation velocity $v$ with exactly $\beta = 4$.

We have seen that $\overline{a}(r)$ is just the effective acceleration which a non-comoving observer measures for objects outside of a certain radius at astronomical distances, where the expansion of the universe matters. Below this radius, $r_0$, this acceleration coincides with Newtonian gravity. It just turns out that in the relevant ranges $\overline{a}(r)$ coincides with the corrections to Newtonian gravity suggested by MOND.

---

[21] $a_0 \sim 1.2 \cdot 10^{-10} \frac{m}{s^2}$.